\documentclass[prb,aps,twocolumn,dcolumn,superscriptaddress,showpacs,amsfonts,amsmath,amssymb,showkeys,floatfix]{revtex4-1}
\usepackage{bm}
\usepackage{pstricks}
\usepackage{setspace}
\usepackage{graphicx}
\begin{document}
\title{Pair correlation function for spin glasses}
\date{\today}
\author{Julio F. Fern\'andez}
\email[E-mail address: ] {jefe@unizar.es}
\affiliation{Departamento de F\'{\i}sica de la Materia Condensada, Universidad de Zaragoza, 50009-Zaragoza, Spain}
\affiliation{Instituto Carlos I de F\'{\i}sica Te\'orica y Computacional,  Universidad de Granada, 18071 Granada, Spain}
\author{Juan J. Alonso}
\affiliation{F\'{\i}sica Aplicada I, Universidad de M\'alaga,
29071-M\'alaga, Spain}
\email[E-mail address: ] {jjalonso@uma.es}

\date{\today}

\begin{abstract}

We extract a pair correlation function (PCF) from probability distributions of the spin-overlap parameter $q$.  The
distributions come from Monte Carlo simulations. 
A measure, $w$, of the thermal fluctuations of magnetic patterns follows from the PCFs. 
We also obtain rms  deviations (over different system samples)  $\delta p$ away from  average probabilities for $q$.
For the linear system-sizes $L$ we have studied, (i) $w$ and $\delta p$ are
 independent of $L$ in the Edwards-Anderson model but scale as $1/L$ and $\sqrt{L}$, respectively, in the Sherrington-Kirkpatrik model.

\end{abstract}

\maketitle

\emph{Introduction}---{After decades of much work, the nature of the spin-glass (SG) phase  is still unclear. } A SG phase is found at low temperature in magnetic systems which are both \emph{quench-disordered} and \emph{frustrated }.\cite {libro,s3}
Random spin positions  as well as random spin-spin couplings are sources of quenched disorder. 
When competition arises, because  not all spin-spin coupling energies can be minimized simultaneously, 
a system is said to be  \emph{frustrated}. \cite{frustration} Fixed disorder and built in competition are the two essential ingredients of \emph{complex} systems.\cite{defc}

The Sherrington-Kirkpatrick (SK) spin-glass model,\cite{SK} in which each spin-spin coupling is assigned at random, without regard to spin-spin distance, is quench-disordered and frustrated. Its exact solution \cite{solution,solution2,solution3,libro} implies 
different random number seeds (which fully specify 
all  couplings) can give rise, in thermal equilibrium,  to magnetic patterns (MPs) which are macroscopically different. \cite{seedD} Diversity of macroscopic observable magnitudes arising from random arrangements of microscopic constituents is the hallmark of complexity.  \cite{complexity,complexity1}
  
However, no consensus has yet been reached on whether 
the \emph{macroscopic} limit of the Edwards-Anderson\cite{EA} (EA), in which only nearest-neighbor spins interact,  (i) follows closely the SK model,\cite{opposite} (ii) deviates from SK model behavior but nevertheless shows  some diversity, \cite{halfway} or (iii) fits a radically different picture, the droplet scenario, \cite{droplet,bokil} in which  SGs with up-down symmetry can only be found in one of two macroscopic spin configurations which are related by global spin inversion. 
Thus, according to the droplet theory, the two necessary ingredients (quenched disorder and frustration) for complexity would become unable to generate diversity in the macroscopic limit of EA systems.

Complexity would make
SGs rather exceptional among the many-particle systems of statistical physics.  On the other hand, they would share this property with systems one finds elsewhere, such as in the life sciences, \cite{fraun} information systems, \cite{is}  optimization problems, \cite{kirk} and  finance. \cite{finance} 
In everyone of these fields complexity and diversity are the rule rather than the exception. Thus,  addressing these issues 
in SGs with standard methods of statistical physics can lead to insight into other seemingly disconnected areas of research.

The basic tool for the characterization of the SG state is  
the spin overlap $q$ between two system states. \cite{libro} To define it, let $\sigma_i^{(1)}$ be the spin of system state $1$ at site $i$, and similarly for $2$. Then,
 $q\equiv N^{-1}\Sigma_i \sigma_i^{(1)}  \sigma_i^{(2)}$, i.e., $q$ is the average (over all sites) spin alignment between states $1$ and $2$. One usually lets states $1$ and $2$
be either (i) of a given time evolution of a given specimen at two widely different times, \cite{apart} or (ii) of two independent ($1$ and $2$) time evolutions of the same specimen.

In macroscopic SK systems,  the probability density (PD), $p(q)$, averaged over all realizations of quenched disorder (RQD), fulfills \cite{libro,solution3}, (i) $p(q)\propto f(q)+\delta (q-q_m)$ for $q\geq 0$, where $f(q)$ is a smooth function of $q$ for $q<q_m$ and  $f(q)=0$ for $q\geq q_m$, and (ii) $p(q)=p(-q)$ if no magnetic field is applied (which we will assume throughout). 

The fact that $f(q)\neq 0$  in the SK model implies the existence of "odd" MPs (in addition to a pair of ordinary MPs that one expects
to observe in all magnetic systems), whence complexity follows. Understandably, attempts at discerning  between the macroscopic behaviors of the SK and EA models have focused on $p(q)$.

Specific system samples are interesting to examine.
At least for \emph{finite}-size EA  systems, seed-dependent spin configurations do appear, much as in the SK model,  \cite{asp}, in thermal equilibrium. 
This is illustrated in Refs. [\onlinecite{asp,pi}], where plots of $p_{\cal J}(q)$ vs $q$ are shown for two different ${\cal J}$ sets of spin-spin couplings.  Whereas some portions of $p_{\cal J}(q)$ differ drastically from sample to sample, the portions for larger values of $q$ are alike and all of them peak near $q=q_m$. We let \emph{self-overlap} spikes (SOS) stand for spikes centered near $q_m$.
Because they are all centered near the same position,
the average of SOS (one for each RQD) over all RQD gives rise to the large peak at $q=q_m$. 
Spikes centered on smaller $q$ values, which vary randomly with different RQD, come from spin overlaps between states that belong to different basins of attraction. Accordingly, we refer to them 
as \emph{cross-overlap} spikes (COS). 

Interesting as it might be, statistical information on spike behavior has not, as far as we know,  been available.
Very little information on COS  follows from the (average) behavior of $p(q)$. 
Cross-over spike statistics would enrich our picture of the SG state, somewhat as the pair correlation function does for the physics of liquids.

We aim to show how COS in $p_{\cal J}(q)$, of which the locations and shapes vary randomly over different RQD, can be added in a coherent fashion, in order to obtain a pair correlation function, $g(q\mid 0,Q)$, which is an average (in a sense which is defined below)  over all RQD of all COS in the $0<q<Q$ range. 
We also  (numerically) calculate the width of $g(q\mid 0,Q)$, which is a measure of MP thermal fluctuations, for $Q=1/2$.
The results we obtain for low temperature ($T$) point to the following behavior: (i) $g(q\mid 0,1/2)$ closely follows  a L\'evy-flight like distribution, \cite{levy} 
(ii) the  pair correlation functions that follow from SOS  and COS are roughly equal, that is $g(q\mid 0,1/2)\approx g(q\mid 1/2,1)$,
(iii) the width of $g(q\mid 0,1/2)$ varies little, if at all, with linear system-size $L$ (scales as $1/L$) in the EA 
(SK) model. Thus, different ranges of spin-spin interaction give rise to 
qualitative differences between the complex behavior of spin glasses.

\emph{Models}---{We study the SK and EA models. In both of them, a $\sigma_i=\pm 1$  spin is located at each $i$-th site of a simple cubic lattice of $N=L^3$ sites.} The interaction energy between a pair of spins at sites $i$ and $j$ is given by $J_{ij}\sigma_i\sigma_j$. We let $J_{ij}=\pm 1/\sqrt{N}$ randomly, without bias, for all $ij$ site pairs in the SK model. The transition temperature $T_{sg}$ between the paramagnetic and SG phase is $T_{sg}=1$. \cite{libro,SK}
For the EA model,  $J_{ij}=0$ unless $ij$ are nearest-neighbor pairs, and we draw
each nearest-neighbor bond $J_{ij}$  independently from unbiased Gaussian distributions of unit variance. 
Then, $T_{sg}\simeq 0.95$. \cite{TcEA}

We let $\langle  u_{\cal J}\rangle_{\cal J}$ stand for
the average of a  thermal equilibrium quantity $u_{\cal J}$
over a number $N_s$ of different  sets of random bonds $\{{\cal J}\}$.

\emph{Pair correlation function}---Aiming for statistical information
on COS at low temperature, we let 
\begin{equation}
G_{\cal J}(q\mid Q_1,Q_2)=\int_{Q_1+h(-q)}^{Q_2-h(q)} dq_1\; p_{\cal J}(q_1)  p_{\cal J}(q_1+q)  ,
\label{defg}
\end{equation}
where  $h(q)=0$ if $q<0$ and $h(q)=q$ if $q\geq 0$. Clearly, $(\Delta Q-\mid q \mid )^{-1}G_{\cal J}(q \mid Q_1,Q_2)$, where $\Delta Q=Q_2-Q_1$,
is the average of $p_{\cal J}(q_1)  p_{\cal J}(q_1+q)$ over the $({Q_1+h(-q)},{Q_2-h(q)})$ domain. 
We term
${G}(q\mid Q_1,Q_2)  \equiv \langle {G}_{\cal J} (q\mid Q_1,Q_2)  \rangle_{\cal J}$  \emph{pair correlation function}.

The integral in Eq. (\ref{defg}) is as for
 the PD to be at $-q$ after a 
 two-step random walk  which starts at the origin, in which the \emph{length} of both steps is identically distributed, but they are taken in opposite directions.
 Note that  ${G}_{\cal J} (q\mid Q_1,Q_2)$ (i) peaks at $q=0$,  \cite{fact} (ii)  is even with respect to $q=0$, since $p_{\cal J}(q)=p_{\cal J}(-q)$, and (iii) is somewhat broader than $p_{\cal J}(q)$ (from the theory of random walks). 

The operation defined in Eq. (\ref{defg}) clearly displaces to $q=0$ any spike in $p_{\cal J}$ within the $(Q_1,Q_2)$ domain. 
Thus, by appropriate choice of $Q_1$ and $Q_2$ values, $G(q\mid Q_1,Q_2)$ enables  one to make comparisons (see below)  \emph{on equal footing} of statistical information on SOS and COS. 

We can also define
$g(q\mid Q_1,Q_2)=BG(q\mid Q_1,Q_2)$, where $B\equiv 1/\int_{Q_1-Q_2}^{Q_2-Q_1}dq\; G(q\mid Q_1,Q_2)$. 
Note that $g(q\mid Q_1,Q_2)$ is  the (conditional) PD that  $q_2-q_1=q$, given that $q_1,q_2\in (Q_1,Q_2)$.
More specifically, 
assume  all  realizations of identical pairs of quenched disordered systems
are evolving independently in equilibrium. Then, $g(q\mid Q_1,Q_2)$ is the PD for $q=q_2-q_1$, 
in the set of  pairs of identical systems in which $q_1,q_2\in (Q_1,Q_2)$ are observed
at infinitely far apart times.

We define  widths of two correlation functions. For a distribution function $F(x)$ such that $ \int_{-\infty}^{\infty} dx\; xF(x)=0$, it makes sense to define a width $\delta x$ by $\delta x F(0)\equiv \int_{-\infty}^{\infty} dx\; F(x)$. Since $g(q\mid Q_1,Q_2)$ is normalized, we let
\begin{equation}
w(Q_1,Q_2)=1/g(0 \mid Q_1,Q_2).
\label{w}
\end{equation}
For short, we let $w_{+}\equiv w(0,1/2)$, $w_{-}\equiv w(1/2,1)$.
For $T\lesssim 0.3$, $w_{-}$ and $w_{+}$ are widths for COS and SOS, respectively.
Furthermore, note (i) $g(q\mid Q_1,Q_2)\le  g(0\mid Q_1,Q_2)$ implies $w_{-},w_{+}\leq 1/2$, (ii) that we can think of $w_{-}$  as an intrinsic width of  $g(q\mid 0,1/2)$, if $w_{-}\ll 1/2$, and similarly for $w_{+}$. 
We also define half-widths at half-maxima $\Gamma_{-}$ and $\Gamma_{+}$ by $2g(\Gamma_{-} \mid 0,1/2)=g(0\mid 0,1/2)$ and $2g(\Gamma_{+} \mid 1/2,1)=g(0\mid 1/2,1)$.  

\begin{table}\footnotesize
\begin{center}
\caption{The number of samples $N_s$ and the time $\tau_s$  taken  for equilibration as well as  for  subsequent averaging is given in thousands of MC sweeps. Acceptance rates for system configuration exchanges at all $T$ are larger than or approximately equal to $A$. $\Delta T$ is the temperature spacing between systems in the tempered MC setup.}
\vspace{0.5 cm}
\begin{tabular}{|c| r  r r  |r r r r |}
\hline
 & & SK&  & &  &  EA    &  \\
\hline
L & 4 & 6 & 8  & 4& 6 &8 & 10 \\
\hline
$\tau_s$ & $50$ & $ 50$ & $100$ &$10$ &$10^2$ &$10^3$ & $10^4$ \\
$N_s$ & $20$    & $10$ & $10$ & $30$ &$40$ &$30$  & $5$ \\
$\Delta T$ & $0.05$  & $0.05$ & $0.04$ & $0.1$ & $0.1$ & $0.05$ & $0.04$ \\
$A$ & $0.7$  & $0.5$ &  $0.4$ & $0.7$ &  $0.4$ & $0.5$  & $0.45$  \\
\hline
\end{tabular}
\end{center}
\end{table}

\emph{Method}---All numerical results given below follow from parallel tempered Monte Carlo (MC) simulations. \cite{tmc,tmc3,paja} We give all times in terms of MC sweeps.

All pairs of systems start running from independent random spin configurations. 
Each system pair is then 
allowed to come, in time $\tau_s$, to equilibrium with each reservoir of a string of  them at $T$, $T+\Delta T$, $T+2\Delta T\;\ldots$,  before readings of $q$ values are taken over an additional $\tau_s$ time span.  From many such readings,
the thermal equilibrium probability $p_{\cal J}(q)$, for a given RQD, is obtained for each temperature.

Relevant parameters for the simulations  are in Table I.

\begin{figure}[!t]
\begin{center}
\includegraphics*[width=80mm]{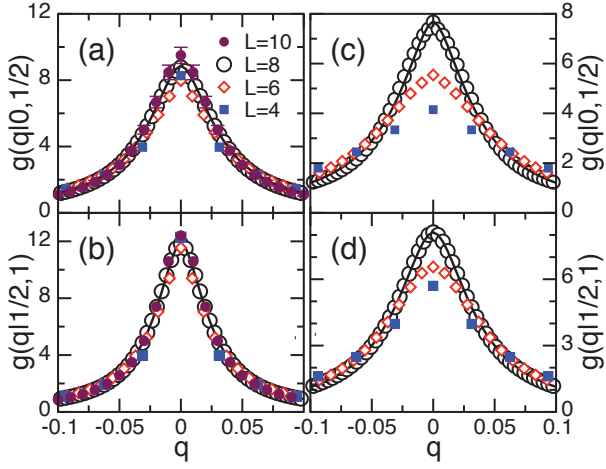}
\caption{(a) Plots of $g(q\mid 0, 1/2)$ vs $q$ for EA systems at $T=0.2$, with the values of $L$ shown. The full line is for a fit to the $L=8$ data points with 
$\Gamma^{\gamma}/[w(\Gamma^{\gamma} +\mid q \mid^{\gamma} )]$.  (b) Same as for (a) but for $g(q\mid 1/2, 1)$. 
(c) Same as for (a) but for the SK model. (d) Same as for (a) but for $g(q\mid 1/2, 1)$ for the SK model. Values of $w$ and $\Gamma$ are given in Fig. \ref{tres} for all $T$ and $L$ above, in both the EA and SK models. Values for $\gamma$ are given in the text.}
\label{spikes}
\end{center}
\end{figure}

\emph{Results for the pair correlation function}---Data points for the pair correlation function are shown in  
Fig. \ref{spikes} for  $T=0.2$. The two graphs on the left hand side (right hand side) are for the EA (SK) model.
The two  top (bottom) graphs are
for $Q_1,Q_2=0,1/2$ ($Q_1,Q_2=1/2,1$), which, since for  $T\lesssim 0.3$, are for COS (SOS).
While neither COS nor SOS exhibit significant size dependence in the EA model, they clearly do so in the SK model. We return 
to this point below.

All curves shown in Fig. \ref{spikes}  are rather pointed at the top. This is in contrast with the well known curves for $p(q)$ in the neighborhood of $q=q_m$ for  finite SK systems \cite{pqm} [but see Ref. \onlinecite{asp,pi} for some $p_{\cal J}(q)$].
This is because values of $q_m$ vary over different RQD by amounts which, at least for $L=8$ and $T\lesssim 0.3$, are roughly equal to $\Gamma_{+}$ for the SK model. 
Thus, averaging $p_{\cal J}(q)$ over all RQD gives rise to a rounded $p(q)$ while $g(q\mid 1/2,1)$, being a \emph{coherent} like superposition of spikes over different RQD, reveals their individual shapes.

\emph{Widths}---We note that if $z=x_1+x_2$, and 
$x_1$ and $x_2$ are drawn from the above distribution, then, within a few per cent,  in obvious notation, $\Gamma_z\simeq \Gamma_x 2/\sqrt{\gamma -1}$ if $1.15\lesssim \gamma \leq 2$.  Thus, $g(q\mid Q_1,Q_2)$ is
approximately $2/\sqrt{\gamma -1}$ times wider  than  spikes in the $ (Q_1,Q_2)$ domain.

\begin{figure}[!t]
\begin{center}
\includegraphics*[width=80mm]{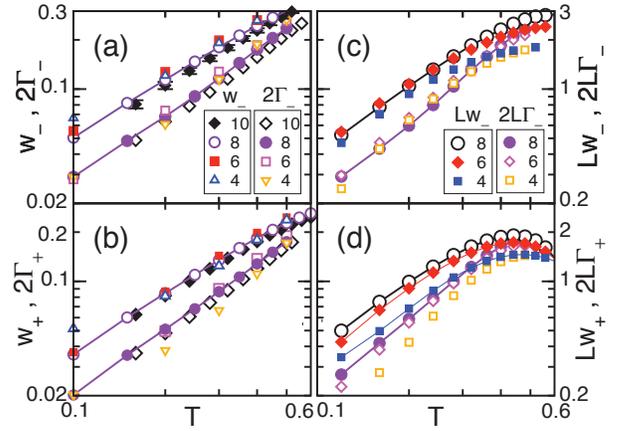}
\caption{(a) Log-log plots of $w_{-}$ and $2\Gamma_{-}$ vs $T$, for EA systems of 
$L=4, 6, 8$ and $L=10$, as shown. Error bars for $L=10$ are shown, but for smaller $L$ values they are hidden by symbols.
(b) Same as in (a) but for $w_{+}$ and $2\Gamma_{+}$.
(c) Same as in (a) but  for  SK systems of $L=4, 6$ and $L=8$, as shown.
(d) Same as in (c) but for $Lw_{+}$ and $2L\Gamma_{+}$.}
\label{tres}
\end{center}
\end{figure}

Good fits to 
all plots in Fig. \ref{spikes} are provided by $\Gamma^{\gamma}/[w(\Gamma^{\gamma} +\mid q \mid^{\gamma} )]$,
which closely follows a L\'evy flight distribution \cite{levy} for $1<\gamma \leq 2$.
Fits  to the $L=8$ data are shown in Figs. \ref{spikes}(a), \ref{spikes}(b), \ref{spikes}(c) and \ref{spikes}(d).
For both the EA and SK models,
$\gamma \simeq2(1 -1/L)$
for all $T\lesssim 0.3$.
Values for $w$ and $\Gamma$ are given in all panels of Fig. \ref{tres}.  As in Fig. \ref{spikes}, the two graphs on the left hand side (right hand side) are for the EA (SK) model.
The two  top (bottom) graphs are
for $Q_1,Q_2=0,1/2$ ($Q_1,Q_2=1/2,1$), which for  $T\lesssim 0.4$ are for COS (SOS).

Widths $w_{-}$ and  $w_{+}$ appear in Figs. \ref{tres}a and \ref{tres}b to be size independent for $0<T\lesssim 0.3$. 
This points to  finite widths for COS and SOS in the $L\rightarrow\infty$ limit of the EA model at low temperature.
On the other hand, $w_{-},w_{+} \sim 1/L$ for large $L$ in the SK model
seems consistent with the
data points shown in Figs. \ref{tres}c and \ref{tres}d for $w_{-}$ and  $w_{+}$. \cite{con}

\begin{figure}[!t]
\begin{center}
\includegraphics*[width=80mm]{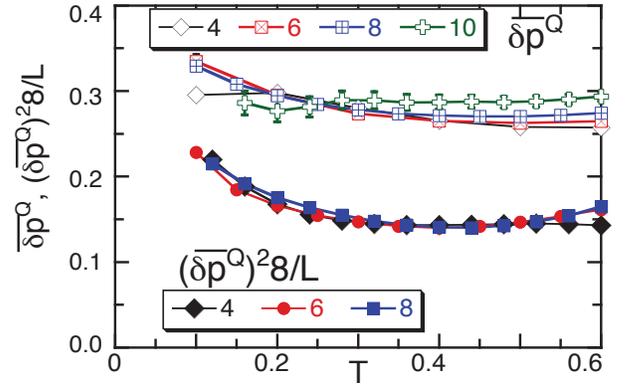}
\caption{Plots of $(\overline{\delta p}^Q)$ and  of $(\overline{\delta p}^Q)^28/L$ vs $T$ for the EA and SK models, respectively, $Q=1/2$ and the values of $L$ shown.
Except for $L=10$, icons cover all error bars.}
\label{ultima}
\end{center}
\end{figure}

\emph{Probability fluctuations over different RQD}---Additional information follows from the (unnormalized) pair correlation function $G(q\mid Q_1,Q_2)$.
For instance, 
\begin{equation}
G(0\mid 0,Q)=Q[(\overline{ p}^Q)^2 + (\overline{\delta p}^Q)^2],
\label{G}
\end{equation}
where $\overline{ p}^Q$ and $(\overline{\delta p}^Q)^2$ are the averages of ${ p}$ and $(\delta p)^2$ over the $0 < \mid q \mid < Q$ range, respectively.

Plots of $(\overline{\delta p}^Q)$ vs $T$ for $Q=1/2$ and various systems sizes of the EA and SK models are shown in Fig. \ref{ultima}. Note that whereas $(\overline{\delta p}^Q)$ scales as $\sim \sqrt{L}$ in the SK model, it appears to be, within statistical errors, independent of $L$ in the EA model.

The amount of fluctuations over different RQD of $X_{\cal J}(Q)$, defined by
$X_{\cal J}(Q)=\int_{-Q}^Qdq\;p_{\cal J}(q)$,
differs qualitatively from $(\overline{\delta p}^Q)^2$ if $Q$ is not too small. This is because,
$\langle  X^2_{\cal J}(Q) \rangle_{\cal J}=w(0,Q)G(0\mid 0,Q)$, as follows from Eq. (\ref{w}). Therefore, our result that $w(0,Q)(\overline{\delta p}^Q)^2$ is independent of $L$ in both the EA and SK models if $Q\gg w(0,Q)$
implies $\langle  X^2_{\cal J}(Q) \rangle_{\cal J}$ remains then bounded in both models  as $L\rightarrow \infty$.

\emph{Conclusions}---We have given a recipe for a coherent addition  of self- and cross-overlap spikes (SOS and COS). The latter are centered on positions that vary randomly with RQD.
In both  the EA and SK models, the correlation functions for SOS and COS turn out to be approximately equal.
The widths $w_{-}$ and $\Gamma_{-}$ (both for COS) give a measure of the thermal fluctuations of magnetic patterns.
They are not too different from $w_{+}$ and $\Gamma_{+}$ (both for SOS), respectively.
Neither $w_{\pm}$ nor $\Gamma_{\pm}$ vary much  with linear system-size $L$ in the EA 
 model but scale approximately as $1/L$ in the SK model.
Their variation with system size at low temperature suggests they vanish in the macroscopic limit of the SK model but remain finite in the EA model.  
Finally, the mean square deviations of $p_{\cal J}$ away from $p(q)$ appear not to vary much  with  $L$ in the EA 
 model but scale approximately as $L$ in the SK model.

The rule we have uncovered --which relates thermal fluctuations  of magnetic patterns as well as probability fluctuations to interaction range-- may well be valid in some broader domain,  beyond the SK and EA models. For one, preliminary (unpublished) work  yields similar results for some \emph{spatially} disordered systems which are geometrically frustrated. Extensions to other fields of complex systems easily come to mind.

\acknowledgments
We are grateful to Larry Falvello for insightful remarks.
We thank SCBI and  LMN, both at Universidad de M\'alaga, for much computer time.
Funding from the Ministerio de Econom\'ia y Competitividad of Spain, through Grant FIS2009-08451,  is gratefully acknowledged.


\begin{thebibliography}{99}
\bibitem{fact} The fact that $G_{\cal J} (0 \mid Q_1,Q_2)\geq G_{\cal J} (q\mid Q_1,Q_2)$, follows easily from the  inequality $(A-B)^2\geq 0$, letting $A^2=B^2=\int dx\;f^2(x)$ and $AB=\int dx\; f(x)f(x+u)$.
\end{thebibliography}
\end{document}